\documentstyle[11pt,aaspp4,flushrt]{article}

\begin{document}

\title
{Internal Kinematics of Blue CFRS Galaxies at $z~\sim~0.6$
\footnote{Based
on observations with the NASA/ESA {\it Hubble Space Telescope}
obtained at the Space Telescope Science Institute, which is operated
by the Association of Universities for Research in Astronomy, Inc.,
under NASA contract NAS 5-26555.}
}

\author
{Gabriela Mall\'en-Ornelas\altaffilmark{2} and  Simon J. Lilly\altaffilmark{2}}
\affil
{Department of Astronomy, University of Toronto, 60 St. George Street,
Toronto, ON M5S~3H8, Canada; mallen@astro.utoronto.ca, lilly@astro.utoronto.ca}

\author
{David Crampton\altaffilmark{2} and David Schade}
\affil
{Dominion Astrophysical Observatory, 5071 W. Saanich Rd., National
Research Council of Canada, Victoria, BC V8X~4M6, Canada; David.Crampton@hia.nrc.ca, David.Schade@hia.nrc.ca}

\altaffiltext{2}
{Visiting Astronomer, Canada-France-Hawaii Telescope, which is
operated by the National Research Council of Canada, the Centre
National de la Recherche Scientifique of France, and the University of
Hawaii.}

\begin{abstract}
   We present the results of a study of the internal kinematics of
luminous starforming galaxies in the $0<z<0.8$ range, with the aim of
investigating the nature of the blue galaxies which cause the largest
changes in the luminosity function at $z\geq0.5$.  New kinematic data
are analysed for a sample of 24 galaxies from the Canada-France
Redshift Survey, most of them with rest-frame $(U-V)_{AB}\leq 1.14.$
Unlike most previous studies, target galaxies were selected {\it
regardless} of size and morphology, from a well-studied
magnitude-limited survey (the CFRS).  Our sample is therefore
representative of the most rapidly changing $1/3$ of the galaxy
population in the $0<z<0.8$ range.  The 15 galaxies at $z>0.45$ have
sizes (from HST images) and velocity widths $\sigma_{v}$ (from
emission lines) similar to those of typical local Irregulars.  This is
consistent with their morphologies and rest-frame colors; however,
these galaxies are as bright as the brightest local Irregulars, and
roughly 2 magnitudes brighter than typical Irregulars known nearby.
We conclude that the increase in the number density of luminous blue
galaxies at $z\geq0.5$ is mainly due to a population of small and
unusually-bright late-type galaxies.

\end{abstract}


\newpage

\section
{INTRODUCTION}

In recent years, redshift surveys have enabled us to construct the
optical luminosity function (LF) for normal field galaxies up to
$z\sim1$, and there is now general agreement that the comoving density
of actively starforming galaxies was higher in the past
(e.g. \cite{sjl95}; \cite{ellis}; \cite{cow}).  The luminosity
function, however, is only a statistical description of the galaxy
population, and additional information on the nature of individual
galaxies is needed in order to understand the causes of any changes
seen in the LF.  

The internal kinematics of a galaxy are closely related to its mass,
and can help place constraints on the kind of galaxy that is being
observed; in particular, different types of galaxies show
characteristic scaling relations of size and velocity width
$\sigma_{v}$, or rotation velocity $V_{rot}$.  In principle, these
scaling relations provide clues about the {\it nature} of a galaxy
independently of any changes in luminosity or morphology that may be
caused by bursts of star formation.

The internal kinematics of distant galaxies can most readily be
studied via optical emission lines, such as H$\alpha$ or the [OII]
3727\AA\ line, and in recent years several kinematic studies of faint
galaxies have been published (e.g. \cite{koo}; Guzm\'an et al.\ 1996,
1997, 1998; \cite{for}; Vogt et al.\ 1996, 1997; \cite{rix};
\cite{phil}; \cite{sim}).  Some of these groups report significant
evolution at even moderate redshifts, while others find little or no
evolution to $z\sim1$.  However, sample selection criteria varied
widely, so these apparent discrepancies could potentially be
reconciled if evolution is not uniform for all types of galaxies.

   We have undertaken a kinematic study of a sample of starforming
galaxies selected from the Canada-France Redshift Survey (CFRS) for
which we have Hubble Space Telescope (HST) images. Our aim was to look
at the general starforming galaxy population at $z\sim0.6$, where
changes in the blue LF are clearly evident. Our sample was selected
{\it regardless of galaxy size or morphology}, and is representative
of the bluest $1/3$ of the galaxies in the CFRS.  

The sample is described in \S 2, the extraction of velocities is
discussed in \S 3, and the results from our study are presented in \S
4.  The main conclusions are summarized in \S 5. Throughout this paper, we
have adopted $H_{0}\!=\!50\,h_{50}$ $\rm km\,s^{-1}Mpc^{-1}$ and
$q_{0}=0.5$.

\section
{THE SAMPLE AND OBSERVATIONS}

New spectroscopic observations were undertaken with the Subarcsecond Imaging
Spectrograph (SIS) at the Canada-France-Hawaii Telescope (CFHT) on
1998 June 21-25, 1996 April 20-23, and 1995 October 27-30.  The SIS is
equipped with an active guiding system, yielding a typical PSF of
0.5-0.7'' FWHM.  Effective slit widths were typically 0.5-0.8'', and
resolutions (FWHM of arc lines) were in the range 1.2-4.1 \AA.

Target galaxies were selected from the subset of the CFRS which has
been imaged by HST (see \cite{brinch} and references therein).  Based
on the HST images, we constructed multi-object masks in which slits
were aligned along the major axis of each galaxy.  Slits were placed
over every available known galaxy {\it regardless} of its size or
morphology, but with preference given to galaxies with strong emission
lines.  Objects with unreliable spectra (e.g. a bad column, too close
to the edge of the slit) were removed from the sample.

   The sample selection is illustrated in Figure 1, which shows a plot
of rest-frame $(U-V)_{AB}$\ vs.\ $z$ for our sample (large symbols)
and other CFRS galaxies (small crosses).  The 30 galaxies for which we
were able to measure the velocity width of their emission lines (large
filled symbols) were generally blue.  We successfully measured
$\sigma_{v}$ for every observed galaxy with rest-frame
$(U-V)_{AB}\leq1.14$ (dotted line), which is roughly the color of a
local Scd galaxy (\cite{cww}).  The typical [OII] 3727\AA\ rest-frame
equivalent width for the $z\sim0.6$ blue galaxies is $\sim$60\,\AA , as
compared to $\sim$40\,\AA\ at $z\sim0.3$. The sample has the same
$I_{AB}<22.5$ magnitude limit as the CFRS; thus, our kinematic sample
represents the bluest $1/3$ of the CFRS galaxies, which is precisely
the population that shows the strongest evolution in the luminosity
function to $z\sim1$ (\cite{sjl95}).

\section {EXTRACTION OF VELOCITIES} 

   We chose the velocity width $\sigma_{v}$ to characterize the
kinematics, since (a) the nature of galaxies in our sample was not
known {\it a priori}, and $\sigma_{v}$ is a general parameter which
applies to both ordered rotation and random motions, (b) many galaxies
at high $z$ were too small to reliably measure rotation curves, and
(c) summing all the light increased the signal-to-noise ratio (S/N) in
our spectra.  Similarly, we chose the half-light radius r$_{1/2}$ to
characterize galaxy sizes, since it is a general parameter that
applies to galaxies with any light profile. Half-light radii were
derived from two-dimensional modeled fits to the HST images, and
account for the effects of the PSF (see section 6 of \cite{sjl98}).

Internal velocities were measured from the [OII] 3727\AA\ doublet for
galaxies at $z>0.45$, and from [OIII] 5007\AA\ or H$\beta$ for those
at $z<0.4$.  Emission lines were fit by Gaussians, using the
fit/deblending option within the {\tt splot} task in IRAF, yielding an
observed velocity width $\sigma_{obs}=$FWHM$/2.355$. Gaussian fits
to calibration and sky lines were used to obtain an instrumental
resolution width $\sigma_{res}$ for each spectrum, and the velocity
width $\sigma_{v}=(\sigma_{obs}^2-\sigma_{res}^2)^{1/2}$ was then
computed for each object.  Error bars reflect the uncertainties of the
Gaussian fits and the effects of poor sampling and non-Gaussian slit
profiles (based on simulations).


A detailed analysis of the potential systematic errors will be
presented in Mall\'en-Ornelas et al. (1999) but the main issues are
addressed here.  One concern is that the emission line used for the
kinematic measurements may be coming from a small region of the
galaxy, and thus yield an underestimate of the true velocity width.
We compared the spatial extent of the emission line on the 2-d
spectrogram and found it was consistent with the size of the continuum
in every case except one.  This discrepant galaxy was one of the three
which showed morphological evidence for a merger, and it indeed had a
very large velocity width; the three galaxies with merger morphology
have been omitted from the analysis below.

Another possible concern is the effect of galaxy inclination on the
measured projected velocities.  The three face-on galaxies in our
sample (axial ratios in the range $0.8\leq{b/a}\leq1.0$) have been
omitted from the velocity analysis, since in addition to the
projection effects on the measured velocity, there is considerable
uncertainty in placing the slit along the kinematic major axis of the
galaxy.  Slit orientation should not have a significant effect on the
rest of the sample, since slits were oriented along the major axis of
each galaxy, as measured from HST images.  Finally, there is little
worry that we have a detection bias against galaxies with large
internal velocities, since we have measured $\sigma_{v}$ for a
complete magnitude-limited sample of galaxies with rest-frame
$(U-V)_{AB}\leq 1.14$.  After the three mergers and the three face-on
galaxies were removed, we were left with a sample of 24 galaxies to be
used in the analysis.

\section {RESULTS}

Galaxies in our sample were examined by comparing their kinematics
($\sigma_{v}$), sizes ($r_{1/2}$) and luminosities ($M_{B}$) with
those of various types of local galaxies.  Figure 2 shows $\sigma_{v}$
vs. $r_{1/2}$ and $\sigma_{v}$ vs. $M_{B}$ for samples of local
galaxies (top 8 panels, small symbols), CFRS galaxies at $z<0.4$
(top 8 panels, large symbols), and CFRS galaxies at
$z>0.45$ (bottom 2 panels).  Data for local bulges are from
Bender et al.\ (1992), and data for HII galaxies are from Telles
\& Terlevich (1997) and Telles, Melnick, \& Terlevich (1997).  Data for
Spiral and Irregular galaxies are from the Third Reference Catalogue
of Bright Galaxies (\cite{rc3}, hereafter RC3).  For this last data
set, we took $\sigma_{v}=W_{20}/2.35$.  As with the CFRS sample,
local face-on Spirals and Irregulars ($0.8\leq{b/a}\leq1.0$) were
omitted, but no other inclination correction was made.

The $z<0.4$ CFRS galaxies were classified morphologically from their
HST images, and are plotted in the corresponding panels.  On the
$\sigma_{v}$ vs. $r_{1/2}$ diagram (left panels), most $z<0.4$ CFRS
galaxies lie within the loci defined by local galaxies of the same
morphological class.  The $\sigma_{v}$ vs. $M_{B}$ plots (right
panels) show that these galaxies have luminosities similar to those of
local galaxies of similar morphological type and kinematics, but those
classified as Late Spirals or Irregulars are generally on the brighter
side of the locus.  However, we will not explicitly translate this
into a change in luminosity (or density) since (a) neither our CFRS
sample nor our local comparison sample are complete for the fainter
magnitudes, so it is impossible to compute a shift in luminosity
between the two populations, and (b) our $z<0.4$ blue sample is quite
small, the velocity widths have large error bars, and many of these
widths were measured on the [OIII] 5007 \AA\ line, which may lead to
poorly understood systematic errors.  Thus, the amount of any
evolution required in the $z<0.4$ population is uncertain.

The $0.45<z<0.8$ CFRS galaxies (with median $z\sim0.6$) have Irregular
or late Spiral morphologies, and are plotted separately in the bottom
two panels.  These galaxies lie in a small region of ($\sigma_{v}$,\
$r_{1/2}$,\ $M_{B}$)-space, with typical values of
$\sigma_{v},\sim60\,$km/s,$\ r_{1/2}\sim3\,h_{50}^{-1}$kpc,\ and
$M_{B}\sim-20.5+5\,$log$\,h_{50}$.  This region is outlined with a
box which is reproduced in the upper 8 panels.  The small range in
$M_{B}$ comes from the fact that the original CFRS was a
magnitude-limited sample.

An object's location in the $\sigma_{v}$ vs. $r_{1/2}$ plane is {\it
in principle} independent of luminosity evolution and can serve to
identify possible local counterparts of the $z\sim0.6$ population.  As
can be seen from the left-hand panels of Figure 2, the sizes and
kinematics of the $z\sim0.6$ CFRS galaxies are similar to those of
typical Irregulars in the RC3. This is consistent with their
morphologies as seen in HST images (\cite{brinch}) and their blue
rest-frame colors.  However, one can see on the $\sigma_{v}$
vs. $M_{AB}(B)$ panel (on the right) that only the brightest local
Irregulars are as bright as the $z\sim0.6$ galaxies, and a typical
local Irregular with similar size and kinematics is $\sim$2 magnitudes
dimmer.  Note that changing the cosmology to one of lower $q_{0}$
would not significantly alter our result (e.g. adopting $q_{0}=0.1$ would make the $z\sim0.6$ galaxies $\sim$12\% larger and
$\sim$0.25 mag brighter).

Thus, when we look at the galaxies with blue colors and $-20\leq{M_{AB}(B)}\leq -21.5$ that are producing the largest changes
in the galaxian luminosity function to $z\sim0.6$ (\cite{sjl95}), we
find galaxies that have the colors, morphologies, sizes, and
kinematics of typical Irregular galaxies in the present-day Universe,
but which are $\sim$2 mag brighter than the average Irregular galaxy
seen nearby.  There are, however, some bright Irregulars nearby
(e.g. NGC~5464, NGC~4194, NGC~3239 and Mrk~330) that have same
$\sigma_{v}$,\ $r_{1/2}$, {\it and} $M_{B}$ as the high-redshift
galaxies.

Determining the average evolutionary change in luminosity ($\Delta
M$) is almost impossible from a magnitude-limited sample such as the
one used here --- and in all other studies --- unless one can claim to
be ``complete'' in some class of galaxy (c.f. the discussion of ``big
disks'' in \cite{sjl98}); otherwise, it is impossible to talk about
the change in luminosity for a given class of galaxies, when we cannot
see the fainter members of the class.  Thus, our size and kinematic
data for blue galaxies at $0.45<z<0.8$ are consistent with either (a) a
luminosity evolution $\Delta M_{B}\sim2$ mag to $z\sim0.6$, or (b) a
density enhancement by a factor of $\sim4$ (based on the LF) and no
change in luminosity, or (c) a combination of both luminosity and
density evolution.

A full comparison with other kinematic data in the literature will be
presented in Mall\'en-Ornelas et al.\ (1999), but it should be noted
here that the amount of luminosity evolution inferred by different
authors strongly depends on the local sample used for comparisons.  In
particular, we stress that the Tully-Fisher relation as defined by
spiral galaxies is not necessarily the most appropriate comparison for
{\it small} galaxies with low masses, as seen at high redshift,
regardless of their luminosities.  Furthermore, as long as one is
sampling a small range of magnitudes, it is impossible to tell whether
what is usually interpreted as luminosity evolution of the observed
galaxies, is not in fact caused by an increase in the density of a
larger population, of which we are only seeing the brightest members.
Finally, one must be cautious in assuming that the sizes of individual
galaxies remain constant between $z\sim0.6$ and the present epoch,
since hierarchical merging scenarios predict that galaxies are still
growing at these redshifts (e.g. \cite{mmw}).

Our results are consistent with the studies of compact galaxies by
Guzm\'an et al.\ (1996, 1997, 1998) and Phillips et al.\ (1997) who
found that small galaxies at $0.4<z<1.4$ had high rates of star
formation, and provided a large contribution to the evolution of the
blue population to $z\sim1$.  Of the 15 blue galaxies in our
$0.45<z<0.8$ sample, 12 meet the Guzm\'an/Phillips size and magnitude
selection criteria, so it is not surprising that we have reached
similar conclusions.  It should be emphasized that our results do not
disagree with those of Vogt et al.\ (1996, 1997), who found little or
no evolution from rotation curves of a sample of large disk galaxies.
The blue population, studied here is made up of {\it small}
galaxies which would not be included in the Vogt et al.\ sample (see
discussion in
\cite{sjl98}).

Although the current generation of kinematic studies is a start,
untangling the physical processes responsible for the evolution of the
galaxy population will ultimately require detailed physical
information (including masses) from large samples of galaxies,
extending so far down the luminosity function that the serious
limitations of magnitude-limited samples are overcome.

\section {SUMMARY}
We have measured the internal kinematics of a sample of blue CFRS galaxies
[rest-frame $(U-V)_{AB}\leq1.14$]\ for which we have HST images, and have
found the following:

1. Luminous blue galaxies at $z\sim0.6$ are generally small, have
late Spiral or Irregular morphology, and have kinematics and sizes
consistent with those of local Irregulars or large HII galaxies, 
but they are as bright as the brightest local Irregulars, and roughly
2 magnitudes brighter than typical Irregulars nearby.

2. The major contribution to the evolution of blue L$^{*}$ galaxies to
$z\sim0.6$ comes from a population of strongly evolving small blue
galaxies of mass and morphology similar to that of present-day
Irregulars.

\acknowledgments

We thank M. Sawicki and S. Courteau for useful suggestions.  This
research has been supported by the North Atlantic Treaty Organization
and the Natural Sciences and Engineering Research Council of Canada.
SJL is a fellow of the Canadian Institute for Advanced Research.

\newpage

\begin{figure}
\plotone{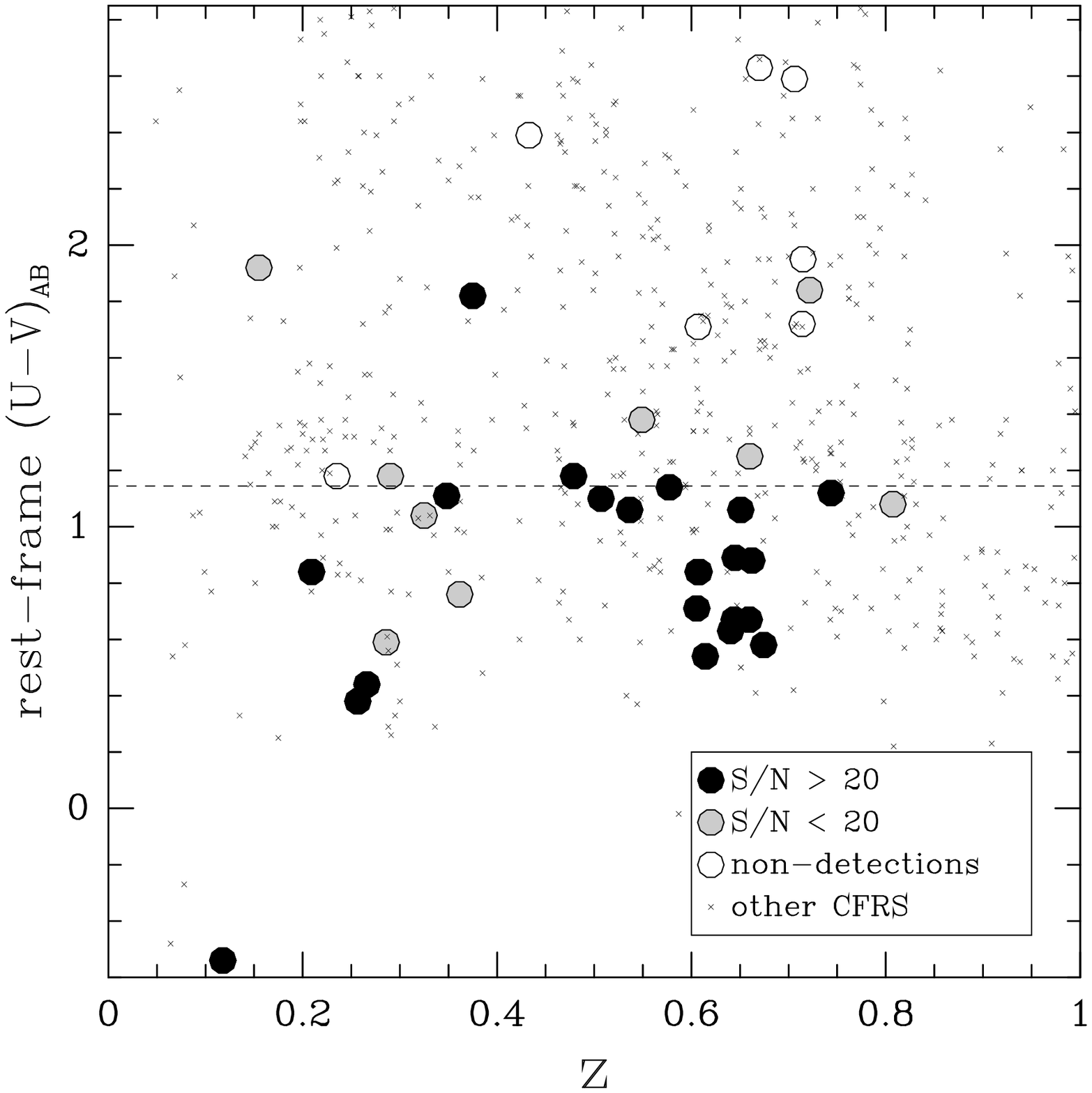}
\caption[gmokinfig1.ps]{Plot of rest-frame $(U-V)_{AB}$ vs. $z$
for the CFRS sample, computed as in Lilly et al.\ (1995).  Galaxies
which were observed in our kinematic study are shown as large symbols,
while the rest of the CFRS galaxies are shown as small crosses.
Objects for which we failed to detect an emission line due to
instrumental reasons have been omitted. Note that the sample shown
above includes the three mergers and three face-on galaxies which were
excluded from the analysis.}
\end{figure}

\newpage

\begin{figure}
\plotone{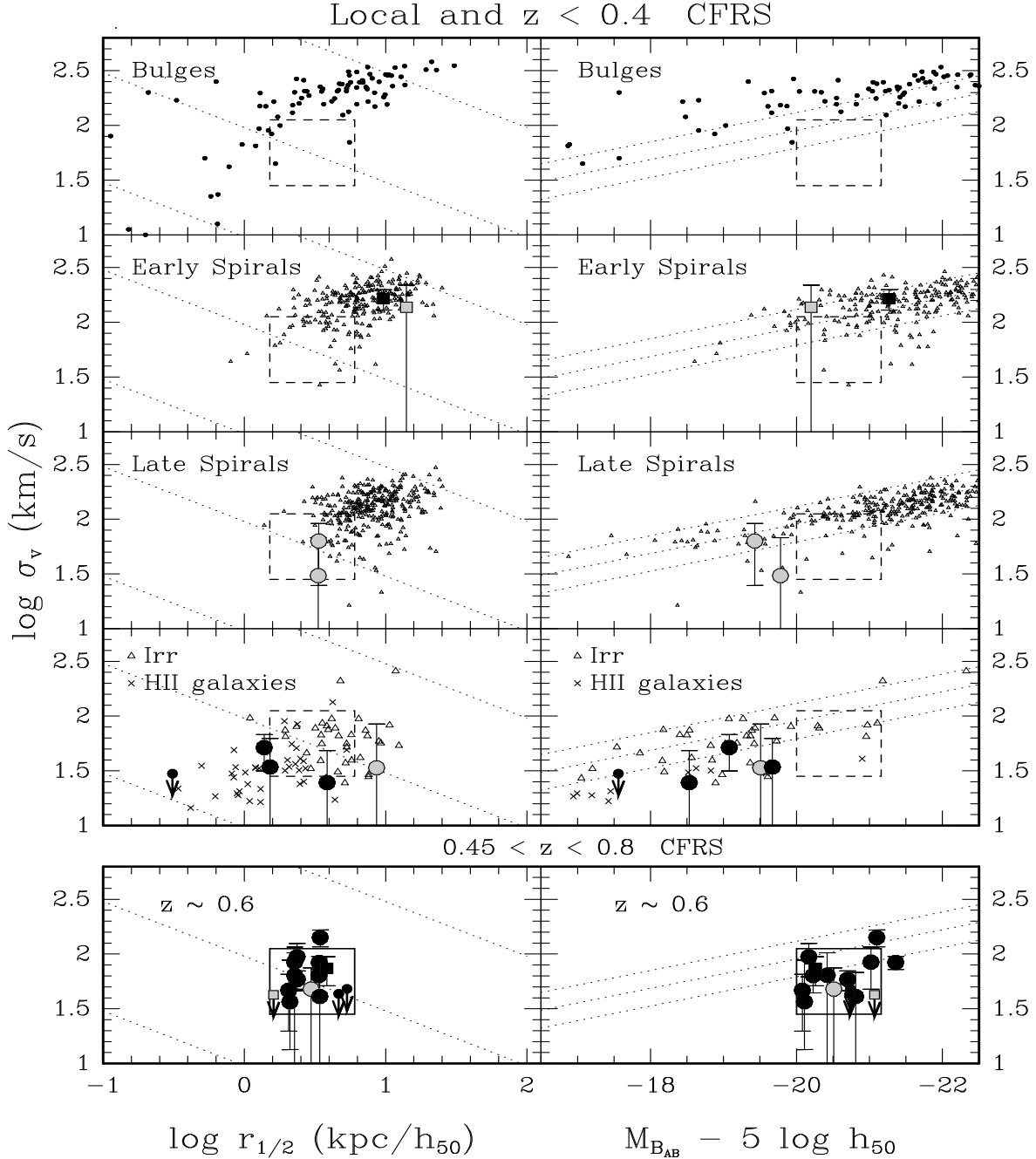}
\caption[gmokinfig2.ps]{\footnotesize Two projections of the ($\sigma_{v}$,
$r_{1/2}$,\ $M_{B}$) space for a sample of local galaxies (top 8
panels, small symbols), for CFRS galaxies at $z<0.4$ (top 8 panels,
large symbols with error bars), and for CFRS galaxies at $z>0.45$
(bottom two panels). Gray symbols represent $S/N<20$ objects;
galaxies with rest-frame $(U-V)_{AB}\leq1.14$ are shown as
circles, and those with rest-frame $(U-V)_{AB}\geq1.14$ are shown
as squares; points with arrows represent 1-$\sigma$ upper-limits. The
dotted lines are fiducials (constant mass lines on the left-hand
panels, and log$\sigma = 0.134 M_{B} + C$ on the right-hand
panels). The dashed box in the upper eight panels indicates the
location of the $z>0.45$ CFRS galaxies as defined from the bottom two
panels.}
\end{figure}

\end{document}